\documentclass[12pt]{article}

\usepackage{scicite}
\usepackage{epsfig,amssymb,amsmath,lscape, color}

\usepackage{times}
\usepackage{caption}

\newcommand{\mnras}{Mon.~Not.~R.~Astron.~Soc.}
\newcommand{\apj}{Astrophys.~J.}
\newcommand{\apjl}{Astrophys.~J.~Lett.}

\newcommand{\nat}{{\it Nature}}
\newcommand{\pasp}{{\it Pub. Astron. Soc. Pac.}}
\newcommand{\physrep}{{\it Phys. Rep.}}
\newcommand{\aap}{{\it Astron. Astrophys.}}

\newcommand{\procspie}{{\it Proceedings of the SPIE}}

\def\arcdeg{\mbox{$^\circ$}}

\def\ldssT{$11,000^{+3400}_{-900}$\,K}
\def\ldssR{$3.3^{+0.3}_{-0.8} \times10^{14}$\,cm}
\def\ldssv{$77,000^{+7000}_{-20,000}$\,km\,s$^{-1}$}
\def\ldssvc{$0.26^{+0.02}_{-0.07}$\,c}

\def\mageT{$9300^{+300}_{-300}$\,K}
\def\mageR{$4.1^{+0.2}_{-0.2} \times10^{14}$\,cm}
\def\magev{$90,000^{+4000}_{-4000}$\,km\,s$^{-1}$}
\def\magevc{$0.30^{+0.01}_{-0.01}$\,c}

\newenvironment{sciabstract}{%
\begin{quote} \bf}
{\end{quote}}

\title{Early Spectra of the Gravitational Wave Source GW170817: \\ 
Evolution of a Neutron Star Merger}

\author
{B.~J.~Shappee,$^{1,2\ast}$ 
J.~D.~Simon,$^{1}$ 
M.~R.~Drout,$^{1}$  
A.~L.~Piro,$^{1}$\\
N.~Morrell,$^{3}$
J.~L.~Prieto,$^{4,5}$
D.~Kasen,$^{6,7}$
T.~W.-S.~Holoien,$^{1}$
J.~A.~Kollmeier,$^{1}$  \\
D.~D.~Kelson,$^{1}$
D.~A.~Coulter,$^{8}$
R.~J.~Foley,$^{8}$
C.~D.~Kilpatrick,$^{8}$\\
M.~R.~Siebert,$^{8}$
B.~F.~Madore,$^{1}$
A.~Murguia-Berthier,$^{8}$
Y.-C.~Pan,$^{8}$ \\
J.~X.~Prochaska,$^{8}$
E.~Ramirez-Ruiz,$^{8,9}$
A.~Rest,$^{10,11}$ 
C.~Adams,$^{12}$\\
K.~Alatalo,$^{1,10}$
E.~Ba\~{n}ados,$^{1}$ 
J.~Baughman,$^{4,13}$
R.~A.~Bernstein,$^{1}$ 
T.~Bitsakis,$^{14}$\\
K.~Boutsia,$^{3}$
J.~R.~Bravo,$^{3}$
F.~Di Mille,$^{3}$
C.~R.~Higgs,$^{15,16}$
A.~P.~Ji,$^{1,11}$\\
G.~Maravelias,$^{17}$
J.~L.~Marshall,$^{18}$ 
V.~M.~Placco,$^{19}$
G.~Prieto,$^{3}$
Z.~Wan$^{20}$\\
\\}

\date{}

\begin{document} 


\baselineskip24pt


\maketitle 

\noindent
\normalsize{$^{1}$ The Observatories of the Carnegie Institution for Science, 813 Santa Barbara St., Pasadena, CA 91101, USA}\\
\normalsize{$^{2}$ Institute for Astronomy, University of Hawai'i, 2680 Woodlawn Drive, Honolulu, HI 96822, USA}\\
\normalsize{$^{3}$ Las Campanas Observatory, Carnegie Observatories, Casilla 601, La Serena, Chile}\\
\normalsize{$^{4}$ N\'ucleo de Astronom\'ia de la Facultad de Ingenier\'ia y Ciencias, Universidad Diego Portales, Av. Ej \'ercito 441, Santiago, Chile}\\
\normalsize{$^{5}$ Millennium Institute of Astrophysics, Santiago, Chile}\\
\normalsize{$^{6}$ Departments of Physics and Astronomy, 366 LeConte Hall, University of California, Berkeley, CA, 94720, USA}\\
\normalsize{$^{7}$ Nuclear Science Division, Lawrence Berkeley National Laboratory, Berkeley, CA 94720, USA}\\
\normalsize{$^{8}$ Department of Astronomy and Astrophysics, University of California, Santa Cruz, CA 95064, USA}\\
\normalsize{$^{9}$Dark Cosmology Center, Niels Bohr Institute, University of Copenhagen, Blegdamsvej 17, 2100 Copenhagen, Denmark}\\
\normalsize{$^{10}$ Space Telescope Science Institute, 3700 San Martin Drive, Baltimore, MD 21218, USA}\\
\normalsize{$^{11}$ Department of Physics and Astronomy, The Johns Hopkins University, 3400 North Charles Street, Baltimore, MD 21218, USA}\\
\normalsize{$^{12}$ Division of Physics, Mathematics, and Astronomy, California Institute of Technology, Pasadena, CA 91125, USA}\\
\normalsize{$^{13}$ Massachusetts Institute of Technology, Cambridge, MA, USA}\\
\normalsize{$^{14}$ Instituto de Radioastronom\'ia y Astrof\'isica, Universidad Nacional Aut\'onoma de M\'exico, C.P. 58190, Morelia, Mexico}\\
\normalsize{$^{15}$University of Victoria, Victoria, B.C., Canada}\\
\normalsize{$^{16}$NRC Herzberg Institute of Astrophysics, 5071 West Saanich Road, Victoria, B.C., V9E 2E7, Canada}\\
\normalsize{$^{17}$ Instituto de F\'isica y Astronom\'ia, Universidad de Valpara\'iso, Valpara\'iso, Chile}\\
\normalsize{$^{18}$George P. and Cynthia Woods Mitchell Institute for Fundamental Physics and Astronomy, and Department of Physics and Astronomy, Texas A\&M University, College Station, TX 77843, USA}\\
\normalsize{$^{19}$ Department of Physics and JINA Center for the Evolution of the Elements, University of Notre Dame, Notre Dame, IN 46556, USA}\\
\normalsize{$^{20}$ Sydney Institute for Astronomy, School of Physics, A28, University of Sydney, NSW 2006, Australia}\\

\normalsize{$^\ast$To whom correspondence should be addressed; E-mail:  shappee@hawaii.edu.}



\baselineskip24pt
\begin{sciabstract}
On 2017 August 17, Swope Supernova Survey 2017a (SSS17a) was discovered as the optical counterpart of the binary neutron star gravitational wave event GW170817.  We report time-series spectroscopy of SSS17a from 11.75 hours until 8.5 days after merger.  Over the first hour of observations the ejecta rapidly expanded and cooled.  Applying blackbody fits to the spectra, we measure the photosphere cooling from \ldssT\ to \mageT, and determine a photospheric velocity of roughly 30\% of the speed of light. The spectra of SSS17a begin displaying broad features after 1.46 days, and evolve qualitatively over each subsequent day, with distinct blue (early-time) and red (late-time) components.  The late-time component is consistent with theoretical models of r-process-enriched neutron star ejecta, whereas the blue component requires high velocity, lanthanide-free material.

\end{sciabstract}


Short gamma-ray bursts (GRBs) have long been hypothesized to be produced by neutron star mergers \cite{Paczynski1986,Eichler1989}, and thus they are the most likely electromagnetic counterparts to the gravitational wave signals from binary neutron star coalescence. Unfortunately, because GRB emission is highly non-isotropic \cite{Fong2015}, in many cases the beam of gamma-rays will not be seen by an observer. This has motivated studies of electromagnetic counterparts that are more isotropic, with one of the most popular cases being the so-called macronovae or kilonovae \cite{Li1998,Kulkarni2005,Metzger2010,Roberts2011,Metzger2017}. These transients would result from the outflow of $\sim0.01$ solar masses of neutron-rich material, ejected from the merging neutron stars at $\gtrsim10$\% of the speed of light.
This neutron-rich material is expected to synthesize heavy elements that power a fast transient peaking at red optical or near-infrared (near-IR) wavelengths via their radioactive decay. Furthermore, the r-process nucleosynthesis in these outflows---named from the capture of neutrons onto lighter seed nuclei on a timescale more rapid than $\beta$-decays \cite{Burbidge1957,Cameron1957}---may explain the origin of half the elements heavier than iron in the periodic table \cite{Qian2007,Arnould2007}. Although a handful of candidate kilonovae following short GRBs have been identified \cite{Tanvir2013,Berger2013,Kasliwal2017}, none have been studied in detail or conclusively confirmed.

The overall heating rate from r-process nucleosynthesis is generally agreed upon and fairly robust with respect to the composition  \cite{Metzger2010,Roberts2011}, but the expected spectroscopic appearance of a kilonova is much less clear. Kilonova spectra depend strongly  on the nuclear yields, neutrino flux, geometric orientations, mass and velocity of the ejecta. The neutron-rich outflow is expected to produce elements in the lanthanide series,  which have a large effect on the emergent radiation due to the opacity generated by their numerous bound-bound electronic transitions \cite{Kasen2013}.  Despite considerable theoretical effort, there is no consensus on the expected spectrum of a kilonova \cite{Kasen2013,Barnes2013,Tanaka2013}. There could be multiple components to the ejecta, with different compositions, opacities, geometries, and velocities \cite{Kasen2015}.  In the absence of observational measurements of kilonovae, especially spectroscopy, this variety of theoretical models remains largely unconstrained.

On 2017 August 17, the Laser Interferometer Gravitational Wave Observatory (LIGO) and Virgo Collaboration (LVC) detected GW170817, a gravitational wave (GW) signal from a binary neutron star merger \cite{GCN21509}.  A contemporaneous and weak short GRB was detected by the {\it Fermi} and INTErnational Gamma-Ray Astrophysics Laboratory (INTEGRAL) telescopes \cite{GCN21506,GCN21507}.  Following the GW trigger, our team, the One-Meter, Two-Hemisphere (1M2H) collaboration, identified an optical counterpart, Swope Supernova Survey 2017a (SSS17a), 10.87 hours after the merger \cite{GCN21529,Coulter2017}.
SSS17a was located in the galaxy NGC~4993 \cite{GCN21529, Coulter2017}, at a distance of 40 megaparsecs (Mpc) \cite{Freedman2001}, which is an order of magnitude closer than previous gravitational wave detections.  This discovery was immediately announced to the LVC, and we began a comprehensive followup campaign that extended for nearly three weeks. A companion paper presents extensive ultraviolet (UV), optical, and near-infrared (near-IR) photometry of SSS17a, following the light curve of SSS17a as it reddened and faded \cite{Drout17}. 

11.75 hours after the GW170817 trigger, we obtained an optical spectrum of SSS17a with the Low Dispersion Survey Spectrograph-3 (LDSS-3) on the Magellan/Clay telescope at the Las Campanas Observatory, Chile. This early spectrum shows a smooth blue continuum extending over the entire optical wavelength range (Figure 1) \cite{GCN21547}. In the next hour we obtained three additional spectra at higher resolution using LDSS-3 and the Magellan Echellette (MagE) spectrograph on the Magellan/Baade telescope before SSS17a set and was no longer observable from Chile. The transient faded measurably at the bluest wavelengths during the short time interval covered by these initial spectra, while the spectra redward of 5000 \AA{} did not evolve in this time span \mbox{(Figure 1).}

Motivated by the fact that the UV--optical spectral energy distribution (SED) observed $3-4$ hours later (at $t=0.67$~days after the merger) is well approximated by a blackbody \cite{Drout17}, we fit blackbody models to the observed rest-frame, dereddened spectra to quantify the very early spectral evolution. The fitting procedure is described in \cite{MM}. The best-fitting model results in a temperature of \ldssT\  and radius of \ldssR\ at $t=0.49$~days.  The listed uncertainties represent 90\%\ confidence intervals.  Although the peak of the blackbody is located at UV wavelengths that we do not observe, the combination of the known luminosity and the spectral slope from $3800-10000$\,\AA\ provides significant constraints on the temperature. For material to reach this radius so quickly requires an expansion velocity of \ldssv, or \ldssvc, where $c$ is the speed of light. In comparison, typical supernovae have bulk velocities of $10,000\,{\rm km\,s^{-1}}$. Even the most energetic supernovae, which are associated with long-duration GRBs, have peak measured photospheric velocities of $\sim 20,000-50,000\,{\rm km\,s^{-1}}$ 2$-$3 days post-explosion \cite{Cano2017}, less than what we infer here for the GW counterpart. While the velocity of these systems may be even higher in the first day post-explosion, early spectra within 24 hours of explosion are not widely available for GRB-supernovae. 

For the MagE spectrum at $t=0.53$~days, we fit a blackbody temperature of \mageT\ and radius of \mageR. The uncertainties on the temperature fits are not normally distributed, but we find that the difference in temperature between the LDSS-3 and MagE spectra is significant at $5\sigma$ confidence.  This drop in temperature over only an hour indicates that the expanding ejecta are cooling rapidly. A similar velocity of \magev{} (\magevc) is inferred from this spectrum.    

On subsequent nights, we acquired optical spectroscopy of SSS17a covering more than a week after the explosion using both Magellan telescopes. The majority of the spectra were obtained with LDSS-3, but we also observed the source with the Inamori Magellan Areal Camera and Spectrograph (IMACS) and the Magellan Inamori Kyocera Echelle (MIKE) spectrograph. Our spectroscopic time series of SSS17a spans from 0.49 to 8.46 days after the GW170817 trigger (Figure 2).  A description of the acquisition and reduction of these spectra and a log of all spectroscopic observations are presented in the Supplementary Material \cite{MM}.  

We searched the higher-resolution MIKE and MagE spectra for absorption or emission lines from the host galaxy, as well as for Na\,I~D absorption from the Milky Way.  Using the strength of the Milky Way Na~D absorption we confirmed the foreground reddening determined with other spatially coarser methods \cite{MM}. Host galaxy features have been detected in all available short GRB spectra \cite{Levesque2010,Tanvir2010b,Cucchiara2013,deUgartePostigo2014}.  However, we did not detect any host galaxy lines, and we place 2-sigma limits on Na~D absorption more than 5 times stricter than the absorption detected in GRB130603B \cite{deUgartePostigo2014, MM}.

After the rapidly cooling blackbody observed on the first night, later spectra show that the appearance of SSS17a changed not just quantitatively, but qualitatively each night. At 1.46~days, a broad feature extends from $\sim5000-7000$\,\AA, and the spectrum declines toward shorter wavelengths.  At 2.49~days the peak wavelength of the emission has shifted to $\sim7500$\,\AA, and the spectrum has a distinct triangular shape.  At 3.46 and 4.51~days the fall-off at blue wavelengths steepens, the peak of the optical emission continues evolving redward, and a new feature that increases with wavelength develops in the near-IR part of the spectrum.  By 7.45~days after the merger the spectrum consists of a smooth red continuum, with very little flux detected below 6500\,\AA.  These data reveal that the photosphere continued expanding and cooling, with its temperature declining by a factor of $\sim$4 within a week \cite{Drout17}.  

The spectral features of SSS17a become more complex over time. After the first night they are not well described by either single blackbody fits or the sum of two blackbodies.  All features in the data are smooth and very broad, $\sim2500$\,\AA{} wide for the peak centered at $\sim$7500\,\AA{} in the 2.49~d and 3.46~d spectra, $\sim2000$\,\AA{} wide for the trough centered at $\sim$9000\,\AA{} in the 2.49~d and 3.46~d spectra, and $>1000$\,\AA{} for the near-IR feature at 3.46~d and 4.51~d. It is not clear from the data whether these features should be interpreted as emission centered at the wavelength of the flux maxima or absorption centered on the flux minima.
In either case, from their width and the Doppler effect we can estimate that the material in SSS17a must be moving at a velocity of \mbox{$\sim0.2-0.3$\,c}, which is consistent with the observed lack of narrow lines.  

The photometric evolution of SSS17a strongly suggests two distinct components to the ejecta \cite{Drout17,Kilpatrick17}.  The spectral evolution, as well as the inability to match the early and late time spectra with a single kilonova model, for which we make direct comparisons below, also favors two components. The largely featureless blue component dominates the initial spectrum but quickly fades within the first few days. After three days the spectrum becomes dominated by a red component, corresponding to cooler temperatures, which fades much more slowly. 

The spectral evolution of SSS17a is unlike known astronomical transients, as can be seen in Figure 3, which compares the Magellan spectra taken at $t=0.49$, 3.46, and 7.45~days after the GW170817 trigger to other classes of transients. Because SSS17a was associated with a short gamma-ray transient \cite{GCN21506,GCN21507}, it is natural to investigate whether the SSS17a spectra are consistent with afterglow emission.  Only three short GRBs (all at redshifts $z>0.3$) have available optical spectroscopy  \cite{Levesque2010,Tanvir2010b,Cucchiara2013,deUgartePostigo2014}, of which only GRB130603B \cite{Cucchiara2013,deUgartePostigo2014} is unambiguously classified as a short GRB. In Figure 3, we show spectra of GRB130603B taken $\sim8$ hours after the GRB \cite{deUgartePostigo2014}, which do not resemble SSS17a.  Unfortunately, there are no spectra of short GRBs reported in the literature at epochs later than 1 day.  We therefore cannot compare our later spectra against other short GRBs at similar epochs. Despite the limited comparison sample, we conclude that the early blue spectrum of SSS17a does not resemble previously detected short GRB spectra, which are likely dominated by afterglows from a jet projected along the line of sight, arguing that we are instead seeing emission that is independent of the GRB. This conclusion is bolstered by the lack of X-ray and radio emission at early times, which rule out a broad-band synchroton spectrum as the primary driver of the optical emission \cite{Murguia-Berthier17}.

In Figure~3, we also compare SSS17a to supernovae and other rapid transients. Type Ia and Type Ibc supernovae develop emission lines with characteristic width/velocities of $10,000\,{\rm km\,s^{-1}}$ (0.03c) and evolve over weeks to months rather than days. Young Type II supernovae can have blue, smooth spectra, not dissimilar to the earliest spectra of SSS17a, but this phase lasts for many days. At later times, they settle to a temperature of $\sim$6000~K for $\sim$100 days determined by hydrogen recombination and show hydrogen absorption lines, both properties very much unlike SSS17a. Early spectra of rapid blue transients \cite{Drout2014} can be as blue as SSS17a, but they maintain their blue color for days whereas SSS17a had become much redder by day 2.49. GRB980425/SN1998bw and GRB030329/SN2003dh, supernovae associated with long GRBs, revealed the spectrum of a Type Ic supernova as the afterglow faded and remained detectable for months \cite{Patat2001,Stanek2003}. However, the supernova features seen in GRB980425 are absent in SSS17a, and SSS17a again evolves and fades much faster, becoming undetectable in weeks. 

Detailed comparison of the spectral features of SSS17a to theoretical models is challenging. Current models use only a small subset of the elements synthesized in the neutron-rich outflows, and even for the elements that are included there is considerable uncertainty in the details of their line transitions \cite{Kasen2013,Tanaka2013}. The features that are found in the optical spectra of current models are composed of many different transitions and cannot be reliably associated with specific elements. We therefore focus on the most robust physical features (such as color, velocity and temporal evolution) in our comparison.

The ejecta from a neutron star merger can have different compositions and velocities depending on their origin. This motivates comparison to a few characteristic models. In Figure 4A, we present theoretical models of 0.1 solar masses of lanthanide-rich material that has been dynamically ejected with a velocity of $0.2c$ \cite{Barnes2013}. Such a model is consistent with the power-law evolution of the bolometric luminosity of SSS17a at times $\gtrsim4$\, days, which is consistent with the expectations for r-process heating \cite{Drout17}, although it is not currently possible to identify the particular r-process elements responsible.  Modest scaling, by a factor $\lesssim 7$,  was required to match the overall luminosity of each epoch, but there are qualitative similarities to the spectra from 4.51 days onward. However, this model alone does not reproduce the rapidly evolving early blue phase.

The material generating the early component is likely lanthanide-free to reproduce the blue emission. Such material can be driven by accretion disk winds \cite{Metzger2014} or dynamical ejection from the neutron star-neutron star merger interface \cite{Wanajo2014}. We consider both cases, with the main difference being the velocity of the material. In Figure 4B, we compare with a disk wind model \cite{Kasen2015}, which although it can account for the blue colors at early times, has a number of problems. These include a luminosity that is much too low (the model spectra have been scaled by over an order of magnitude), velocities  ($\lesssim0.1c$) that are much smaller than those we infer from the temperature evolution, and  absorption features that are not seen in our early smooth spectra. We therefore disfavor a disk wind origin.

For the case of lanthanide-free dynamical ejecta, there are fewer theoretical predictions of spectra available for direct comparison.  We replicate the main features of this scenario with a model composed of fast-moving ($\gtrsim0.2c$) lanthanide-free material \cite{Kasen2017}, shown in Figure 4C.
Unlike the previous scenarios, such a model may explain the spectra blueward of 9000\,\AA{} for times up to 3.46 days.  At later epochs, the observed spectra have large red excesses relative to the model, which could be explained if the red component is obscuring the lanthanide-free material as the ejecta evolve \cite{Kilpatrick17}. Such obscuration could occur because the red dynamical ejecta are more concentrated in equatorial regions, while the lanthanide-free material would be launched perpendicular to the binary midplane. This geometry suggests a viewing angle that is neither edge nor pole on. However, the large velocity ($>0.2c$) we infer for the blue component might make obscuration difficult and it is unclear if certain viewing angles and ejecta distributions can satisfy all the constraints from the spectral evolution we observe.

The detailed spectral, kinematic, and chemical data obtained for SSS17a provide multiple constraints for understanding binary neutron star mergers. We find that existing models related to neutron star ejecta may explain some aspects of the spectral evolution we present, but no single model matches all of the main properties. The ejecta likely have a complicated, three-dimensional structure, with large internal variations in velocity and composition, and this complexity may be required to account for the full time-series spectral data in greater detail. On the other hand, alternative physical mechanisms should not be discounted, especially for the early, hot, thermal emission. The smooth spectrum of the very early optical emission allows us to rule out models, namely disk winds, that would be degenerate with photometry alone. 

\newpage

\vspace{0.2in}
\includegraphics*[width=0.8\textwidth]{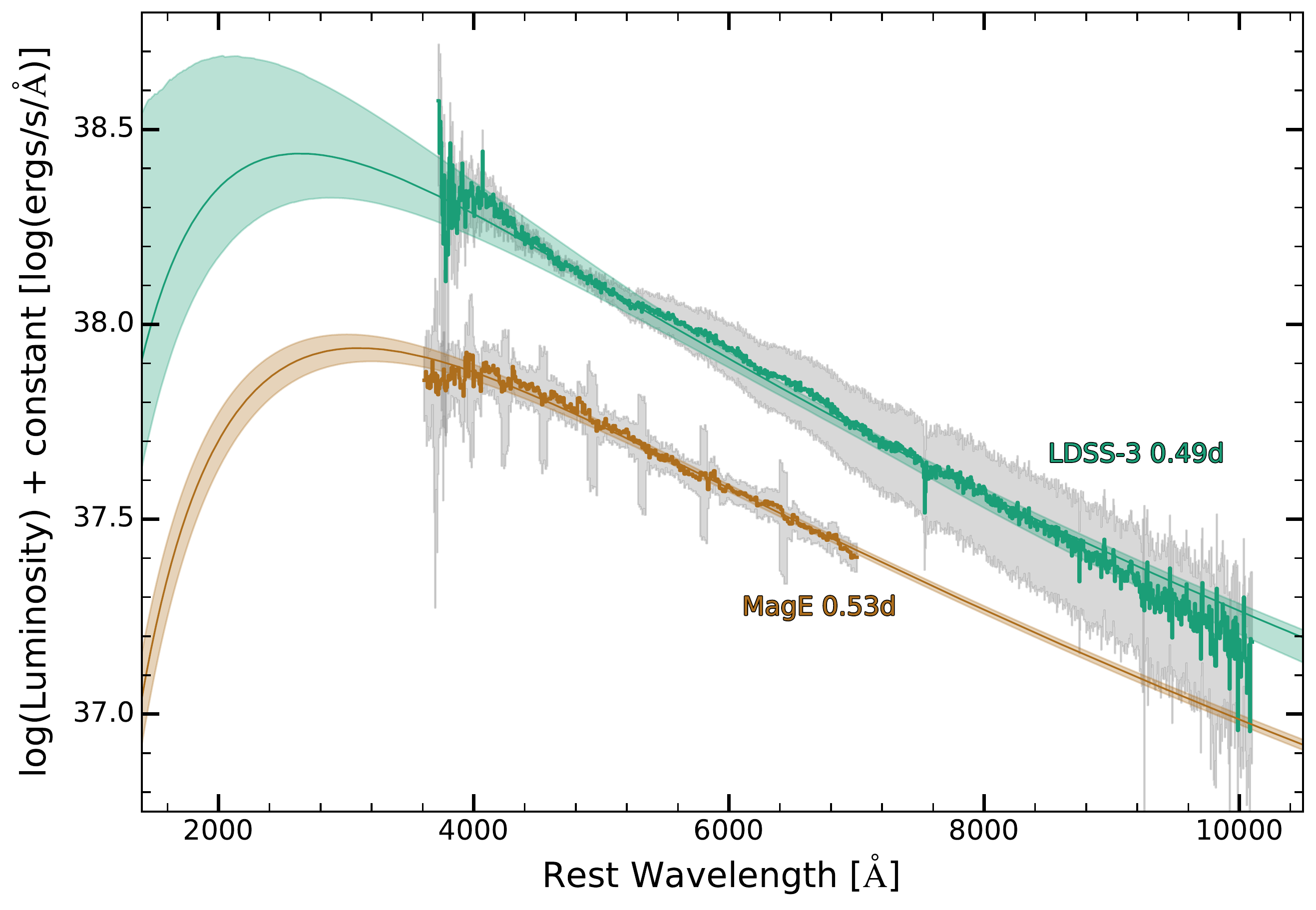}
\newline
\noindent{\bf Figure~1 Early Optical Spectra of SSS17a.}  {Magellan/LDSS-3 and Magellan/MagE spectra of SSS17a acquired 11.75 and 12.75 hours after the LVC trigger, respectively.  The overall slope of the continuum evolved subtly but significantly in this one-hour interval, demonstrating a change in the effective temperature of the source.  Blackbody models and uncertainties are shown by the shaded green (LDSS-3) and brown (MagE) regions. The thick and thin solid lines in the fit regions indicate the median and 90\%\ confidence interval for the fits, respectively. These fits are described in \cite{MM}. The shaded gray outlines surrounding each spectrum indicate the uncertainties on the flux-calibrated spectra. Although the MagE spectrum extends to 10100\,\AA, we only use the data blueward of 7000\,\AA\ for the blackbody fit because of the difficulty of flux-calibrating the data at redder wavelengths where the overlap between adjacent spectral orders is minimal and telluric absorption bands can cover a large fraction of an order.  A vertical offset of $-0.35$~dex has been applied to the MagE spectrum for clarity.} 

\newpage

\includegraphics*[width=0.8\textwidth]{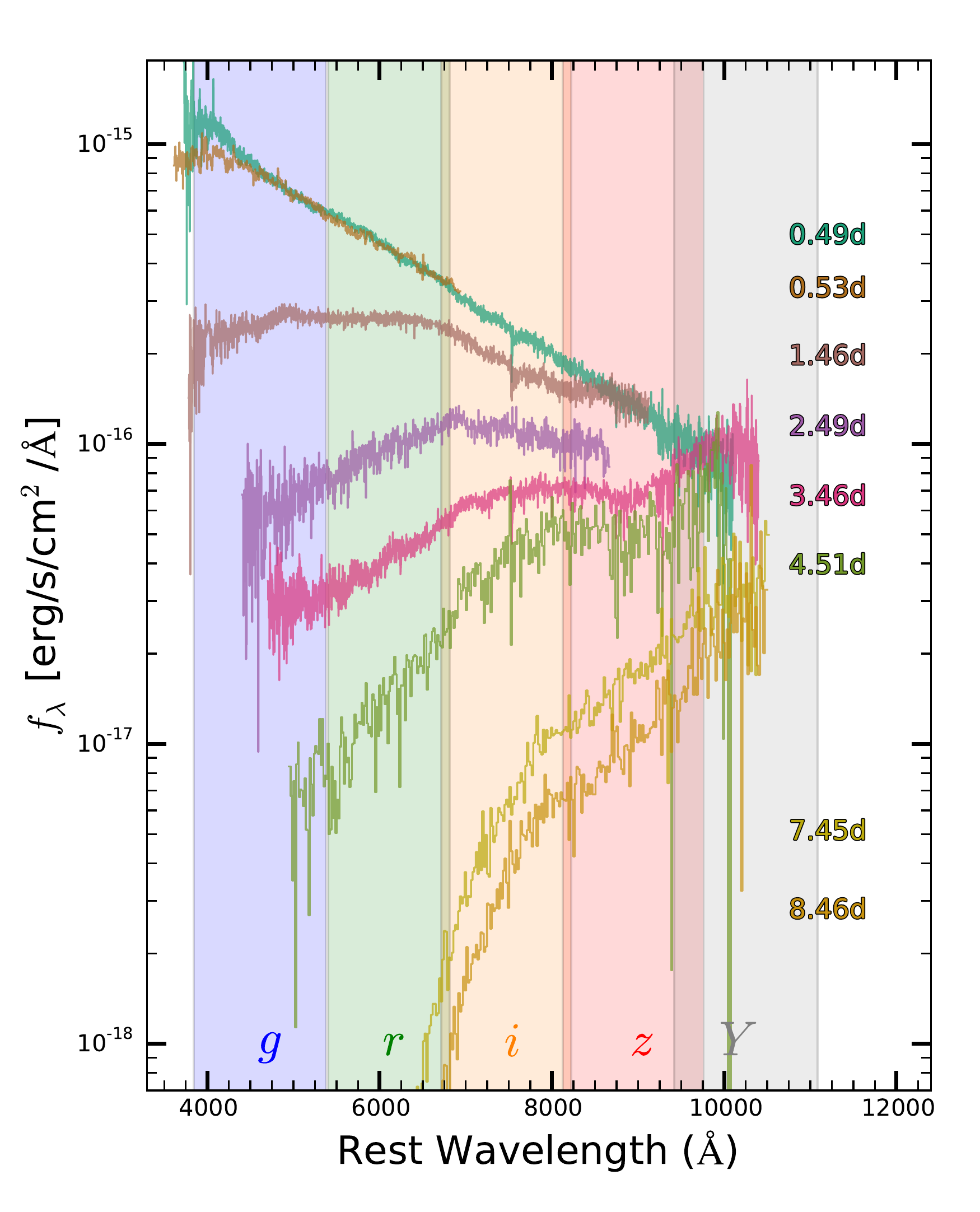}
\newline
\noindent{\bf Figure~2.} {\bf Spectroscopic time series of SSS17a.} {The vertical axis is observed flux ($f_{\lambda}$). Observations began $\sim$ 0.5 days after the merger and were obtained with the LDSS-3, MagE, and IMACS spectrographs on the Magellan telescopes.  These spectra have been calibrated to the photometry of \cite{Coulter2017, Drout17}.  Colored bands indicate the wavelength ranges of the $g,r,i,z$, and $Y$ photometric filters.}

\newpage

\begin{center}
\includegraphics*[width=1.1\textwidth, trim=3cm 0 0 0]{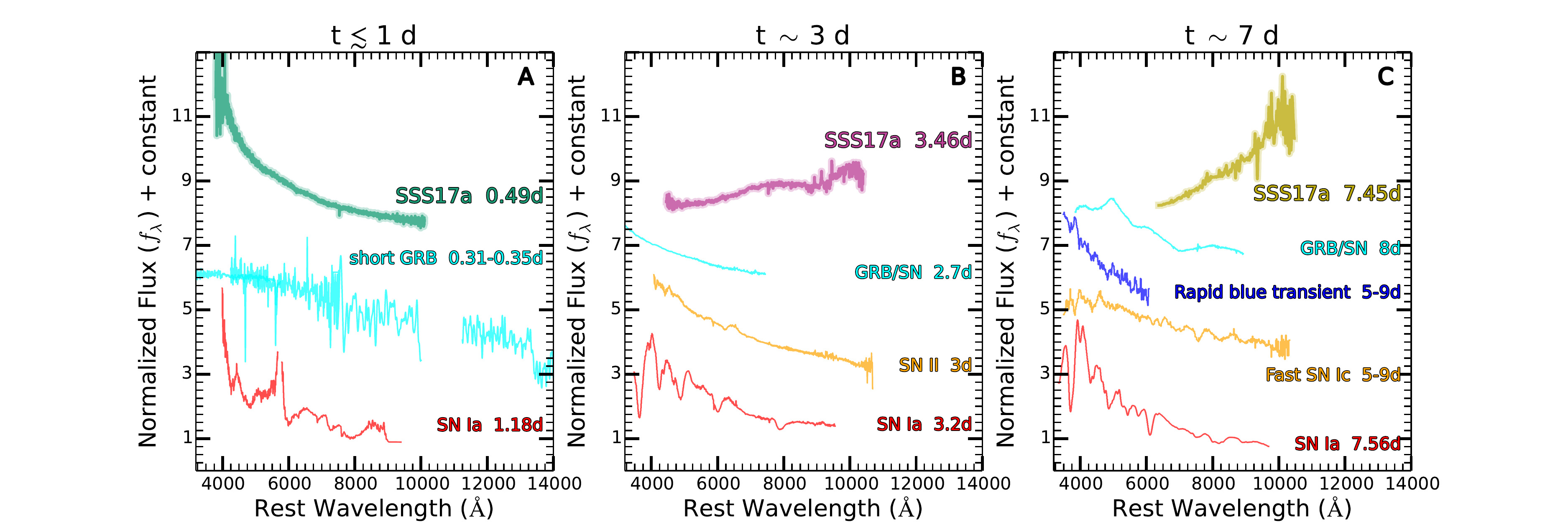}
\end{center}
\noindent{\bf Figure~3.}  {\bf Spectra of SSS17a compared with a broad range of other astronomical transients at several evolutionary phases.}  {While the $\sim0.5$ day spectrum of SSS17a has few features and is potentially an extreme version of some other hot and/or fast transients, it evolves rapidly in comparison. Within 3 days of the LIGO trigger, the optical spectrum of SSS17a is no longer similar to other known transients.  Dates listed are relative to the time of explosion for all objects.  All spectra are divided by their median value and displayed with arbitrary additive offsets for clarity.  {\bf (A)} SSS17a compared to the Type Ia supernova (SN~Ia) SN2011fe \cite{Nugent2011} and the afterglow spectrum of the short gamma-ray burst GRB130603B \cite{deUgartePostigo2014}. Few observations of other transients within 1 day of explosion are available.  {\bf (B)} SSS17a at 3.46~d after explosion compared to the SN~Ia ASASSN-14lp \cite{Shappee2016}, the
Type II supernova SN2006bp \cite{Quimby2007}, and the long gamma-ray burst and its associated afterglow and broad-lined Type Ic supernova GRB030329/SN2003dh \cite{Stanek2003} at similar times relative to explosion.  {\bf (C)} SSS17a at 7.45~d after explosion compared to SN2011fe \cite{Pereira2013}, the rapid blue transient PS1-12bv \cite{Drout2014}, the fast Type Ic supernova SN2005ek \cite{Drout2013}, and the GRB/SN GRB980425/SN1998bw \cite{Patat2001}.
}

\newpage

\vspace{0.2in}
\begin{center}
\includegraphics[width=0.31\textwidth]{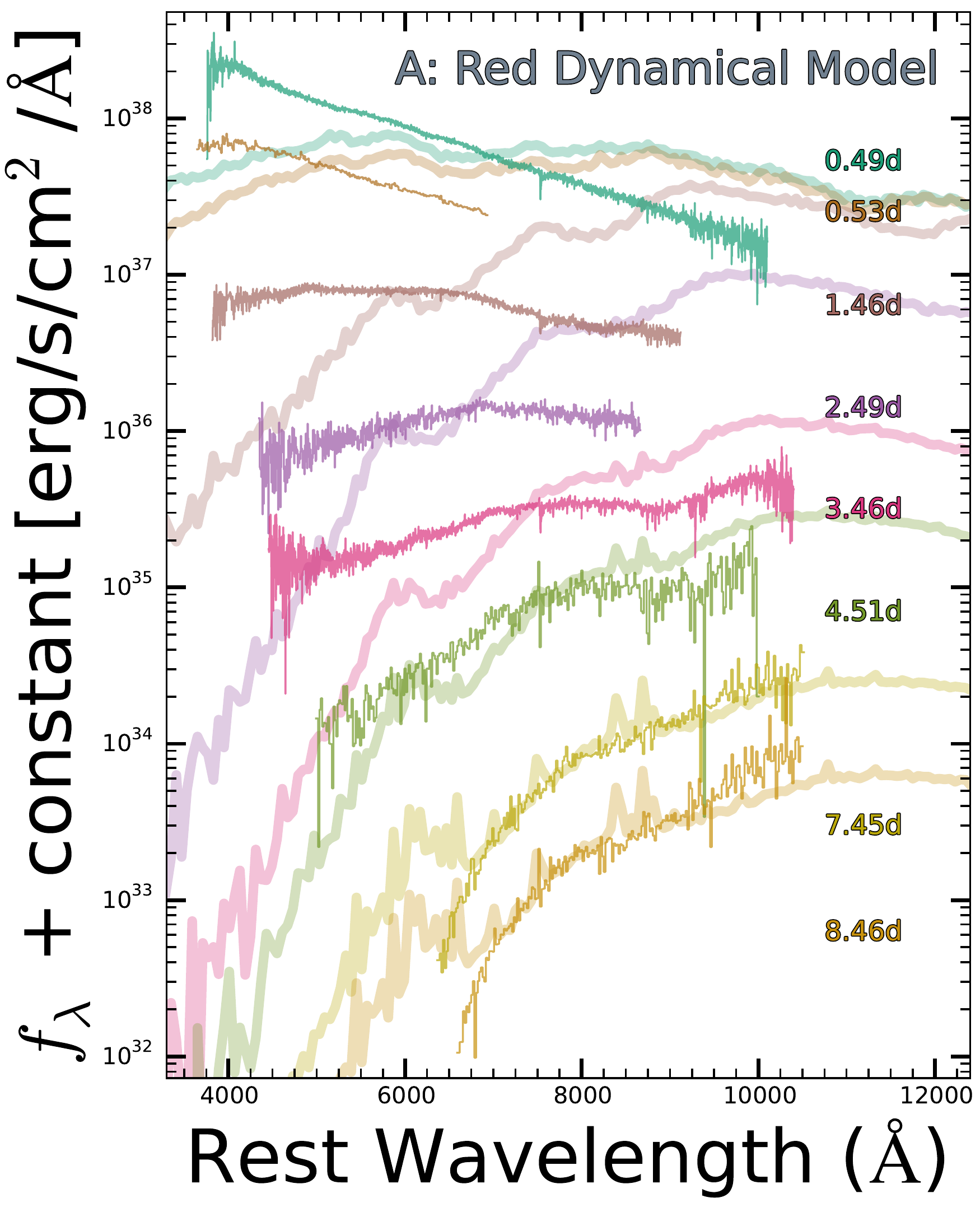}
\includegraphics[width=0.31\textwidth]{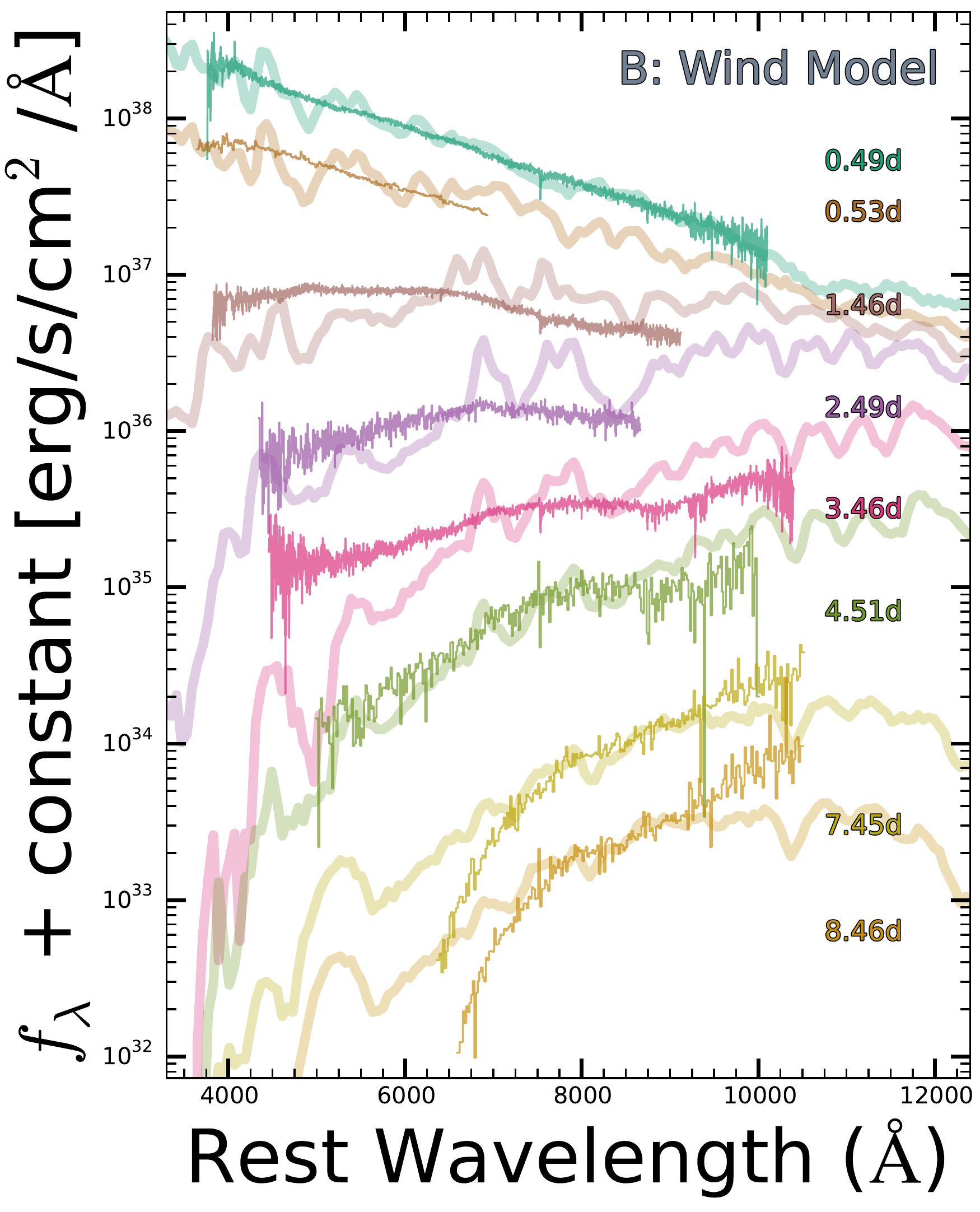}
\includegraphics[width=0.31\textwidth]{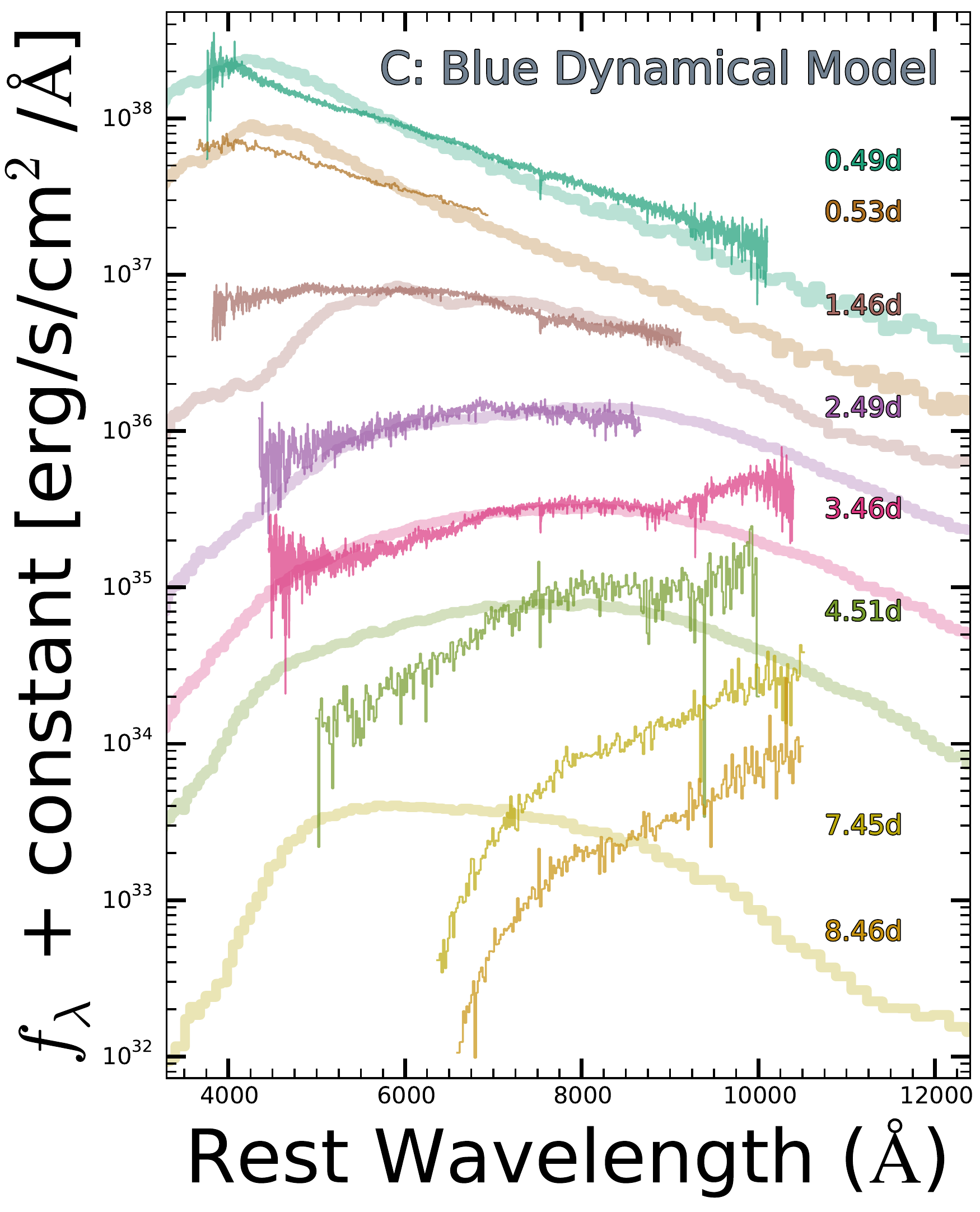}
\end{center}
\noindent{\bf Figure~4. Comparison of SSS17a to theoretical models.}  {The vertical axis is observed flux ($f_{\lambda}$). The models shown are for three possible physical interpretations of SSS17a. While the red kilonova model provides a reasonable likeness to the data at late times, the early time spectra and kinematics require lanthanide-free relativistic material.  No single model shown here or described in the current literature can self-consistently reproduce the full spectroscopic time series of SSS17a.
{\bf (A)}
Lanthanide-rich red kilonova model from a neutron star merger and dynamical ejection \cite{Barnes2013}.  At each epoch, the absolute luminosity is scaled to match the data.
{\bf (B)}
Disk-wind model \cite{Kasen2015} with a neutron star that immediately collapses after the merger.  The absolute luminosity is scaled by a factor of $\gtrsim 10$ at each epoch to match the data.
{\bf (C)}
Lanthanide-poor blue kilonova model from a neutron star merger and dynamical ejection \cite{Kasen2017}.  The model has been crafted to match the observations at early times \cite{Kilpatrick17}.  
}

\bibliographystyle{Science}

\section*{Acknowledgments}

We thank John Mulchaey (Carnegie Observatories director), Leopoldo Infante (Las Campanas Observatory director), and the entire Las Campanas staff for their dedication, professionalism, and excitement, which were critical for obtaining the observations used in this study.

B.J.S., M.R.D., K.A.A., and A.P.J. were supported by NASA through Hubble Fellowships awarded by the Space Telescope Science Institute, which is operated by the Association of Universities for Research in Astronomy, Inc., for NASA, under contract NAS 5-26555. B.J.S. and E.B. were supported by Carnegie-Princeton Fellowships. M.R.D. was supported by a Carnegie-Dunlap Fellowship and acknowledges support from the Dunlap Institute at the University of Toronto. T.W.-S.H. was supported by a Carnegie Fellowship. Support for J.L.P. was in part provided by FONDECYT through the grant 1151445 and by the Ministry of Economy, Development, and Tourism's Millennium Science Initiative through grant IC120009, awarded to The Millennium Institute of Astrophysics. DK is supported in part by a Department of Energy (DOE) Early Career award DE-SC0008067, a DOE Office of Nuclear Physics award DE-SC0017616, and a DOE SciDAC award DE-SC0018297, and by the Director, Office of Energy Research, Office of High Energy and Nuclear Physics, Divisions of Nuclear Physics, of the U.S. Department of Energy under Contract No.DE-AC02-05CH11231.  V.M.P. acknowledges partial support for this work from grant PHY 14- 30152 from the Physics Frontier Center/JINA Center for the Evolution of the Elements (JINA-CEE), awarded by the US National Science Foundation.

The UCSC group was supported in part by NSF grant AST--1518052, the Gordon \& Betty Moore Foundation, the Heising-Simons Foundation, generous donations from many individuals through a UCSC Giving Day grant, and from fellowships from the Alfred P.\ Sloan Foundation (R.J.F), the David and Lucile Packard Foundation (R.J.F.\ and E.R.) and the Niels Bohr Professorship from the DNRF (E.R.).  T.B. acknowledges support from the CONACyT Research Fellowships program. G.M. acknowledges support from CONICYT, Programa de Astronom\'ia/PCI, FONDO ALMA 2014, Proyecto No 31140024.  A.M.B.\ acknowledges support from a UCMEXUS-CONACYT Doctoral Fellowship. C.A. was supported by Caltech through a Summer Undergraduate Research Fellowship (SURF) with funding from the Associates SURF Endowment.

This paper includes data gathered with the 6.5 meter Magellan Telescopes located at Las Campanas Observatory, Chile. Part of this work is based on a comparison to observations of GRB130603B that were obtained from the ESO Science Archive Facility under request number vmplacco308387. 

We thank Antonino Cucchiara for sending us the spectrum of GRB130603B and our thoughts go out to all those in the Virgin Islands impacted by the recent series of hurricanes.  We thank Jennifer van Saders for useful suggestions.  We thank the University of Copenhagen, DARK Cosmology Centre, and the Niels Bohr International Academy for hosting D.A.C., R.J.F., A.M.B., E.R., and M.R.S.\ during a portion of this work.  R.J.F., A.M.B., and E.R.\ were participating in the Kavli Summer Program in Astrophysics, ``Astrophysics with gravitational wave detections.''  This program was supported by the Kavli Foundation, Danish National Research Foundation, the Niels Bohr International Academy, and the Dark Cosmology Centre.

This research has made use of the NASA/IPAC Extragalactic Database (NED), which is operated by the Jet Propulsion Laboratory, California Institute of Technology, under contract with the National Aeronautics and Space Administration.

The data presented in this work and code used to perform the analysis are available at ftp://ftp.obs.carnegiescience.edu/pub/SSS17a. Calibrated rest-wavelength spectra are also made available at WISeREP \cite{Yaron2012} (https://wiserep.weizmann.ac.il/).


\clearpage
\vspace{0.3in}
\noindent {{\bf Supplementary Materials:}\\
{\tt www.sciencemag.org}\\
\noindent Materials and Methods \\
\noindent Figure~S1\\
\noindent Table~S1, S2\\
\noindent References ($48$--$62$)}

\clearpage
\setcounter{page}{1}
\setcounter{figure}{0}    
\renewcommand{\thefigure}{S\arabic{figure}}
\renewcommand{\thetable}{S\arabic{table}}
\renewcommand{\thesection}{S\arabic{section}}
\renewcommand{\theequation}{S\arabic{equation}}

\begin{center}
\title{\LARGE Supplementary Materials for}\\[0.5cm]

\title{{\bf\large The First Spectra of a Gravitational Wave Source: \\ 
Dramatic Evolution of a Neutron Star Merger}}



\author
{B.~J.~Shappee,$^{1,2\ast}$ 
J.~D.~Simon,$^{1}$ 
M.~R.~Drout,$^{1}$  
A.~L.~Piro,$^{1}$\\
N.~Morrell,$^{3}$
J.~L.~Prieto,$^{4,5}$
D.~Kasen,$^{6,7}$
T.~W.-S.~Holoien,$^{1}$
J.~A.~Kollmeier,$^{1}$  \\
D.~D.~Kelson,$^{1}$
D.~A.~Coulter,$^{8}$
R.~J.~Foley,$^{8}$
C.~D.~Kilpatrick,$^{8}$\\
M.~R.~Siebert,$^{8}$
B.~F.~Madore,$^{1}$
A.~Murguia-Berthier,$^{8}$
Y.-C.~Pan,$^{8}$ \\
J.~X.~Prochaska,$^{8}$
E.~Ramirez-Ruiz,$^{8,9}$
A.~Rest,$^{10,11}$ 
C.~Adams,$^{12}$\\
K.~Alatalo,$^{1,10}$
E.~Ba\~{n}ados,$^{1}$ 
J.~Baughman,$^{4,13}$
R.~A.~Bernstein,$^{1}$ 
T.~Bitsakis,$^{14}$\\
K.~Boutsia,$^{3}$
J.~R.~Bravo,$^{3}$
F.~Di Mille,$^{3}$
C.~R.~Higgs,$^{15,16}$
A.~P.~Ji,$^{1,11}$\\
G.~Maravelias,$^{17}$
J.~L.~Marshall,$^{18}$ 
V.~M.~Placco,$^{19}$
G.~Prieto,$^{3}$
Z.~Wan$^{20}$\\
}

\normalsize{Correspondence to: shappee@hawaii.edu.}
\newline
\end{center}

{{\bf This PDF file includes:}\\
\indent \indent \indent Materials and Methods \\
\indent \indent \indent Figure~S1\\
\indent \indent \indent Table~S1, S2\\
\indent \indent \indent References ($48$--$62$)}

\clearpage
\noindent {\bf \LARGE Materials and Methods}

\section{Data Acquisition \& Reductions}

We obtained spectroscopic observations of SSS17a with the Magellan/Clay and Magellan/Baade telescopes beginning 11.75 hours after the neutron star merger, and continued observing SSS17a spectroscopically for 8 days. Below we describe the data acquisition, reduction, and calibration. A log of all spectroscopic observations is given in Table~S1.

\subsection{LDSS-3 Observations}

We observed SSS17a with the Low Dispersion Survey Spectrograph (LDSS-3) on the Magellan/Clay telescope on 2017 Aug. 18-26 (UT).  On the first night we obtained spectra in multiple spectrograph configurations, with the Volume Phase Holographic (VPH)-All, VPH-Blue, and VPH-Red grisms.  The three grisms cover wavelength ranges of $3800-6200$\,\AA, $4250-10000$\,\AA, and $6000-10000$\,\AA\, with resolving powers of $R=1400$, $R=650$, and $R=1400$, respectively.  The later LDSS-3 spectroscopy, when SSS17a was much fainter and redder, employed only the VPH-All or VPH-Red grisms.  All observations were made with a slit width of 1~arcsec. 

We reduced and calibrated the LDSS-3 spectra using IRAF \cite{IRAF} following standard procedures, including bias subtraction, flat-fielding, 1-D spectral extraction, and wavelength calibration by comparison to an arc lamp. Flux calibration and telluric correction was performed using a set of custom IDL scripts \cite{Matheson2008,FCal} based on a spectroscopic standard star observed on the same night.  The statistical uncertainties on the spectra were calculated by standard error propagation.

For the first LDSS-3 spectra of SSS17a, obtained on 2017 Aug. 17-18, the spectrograph slit was oriented $\sim22\arcdeg$ away from parallactic angle.  Given the relatively high airmass (${\rm airmass} \approx 2$) at the time, some flux was lost as a result of differential atmospheric refraction.  A similar offset from parallactic angle was used for LDSS-3 observations on 2017 Aug. 18-19, with smaller flux losses because of the lower airmass.  To correct for this effect, we first calculated the offset of the source from the center of the slit due to the differential refraction, assuming that the target was in the center of the slit at the effective central wavelength of the $g$ filter that was used for target acquisition.  We then measured the atmospheric seeing as a function of wavelength, modeled the source with a two-dimensional Gaussian profile, and computed the fraction of light from the model source that fell in the slit as a function of wavelength.  Finally, the calibrated spectrum of SSS17a was scaled to account for the light that landed outside the slit.  

The magnitude of the differential refraction and seeing corrections depends on the position of the source within the slit.  A source position different than the one we have assumed will result in systematic errors in the scaled spectrum.  Because we do not have a way of measuring this position during the spectroscopic exposure, we quantify the resulting systematic uncertainty by calculating the change in the differential refraction and seeing corrections for source positions offset by 0.1~arcsec in either direction (over a 300\,s exposure the telescope guiding is expected to be at least this accurate) from the nominal position.  At each wavelength, we define the systematic uncertainty to be the average of the changes in the correction factor between the $+0.1$~arcsec and $-0.1$~arcsec offsets.

\subsection{MagE Observations}

We observed SSS17a with the Magellan/MagE spectrograph \cite{Marshall2008} on the nights of 2017 August 17-19.  We used a $0.7~{\rm arcsec} \times 10~{\rm arcsec}$ slit to provide a spectral resolving power of $R=5800$ on August 17-18 and a $1.0~{\rm arcsec} \times 10~{\rm arcsec}$ slit to provide a spectral resolving power of $R=4100$ on August 18-19.  On the first night we obtained a single 322\,s exposure and on the second night we obtained three 1000\,s exposures.  We reduced the MagE spectra using an IDL pipeline based on the techniques described in \cite{Becker2009}, with updates to improve the flux calibration from \cite{Foley2012}.  Observations on both nights were flux calibrated with a standard star spectrum obtained on 2017 Aug. 18-19. 

The spectrum from Aug. 17-18 was taken with the slit oriented at the parallactic angle, so the effects of differential atmospheric refraction should be negligible despite the high airmass (3.0) at the time of observation.  The seeing measured from the spatial profile of the source spectrum varied as a function of wavelength, from 0.9~arcsec at the red end of the spectrum to 1.3~arcsec in the blue.  We corrected for wavelength-dependent loss of light from this seeing variation by modeling the source with a two-dimensional Gaussian profile, calculating the fraction of the flux that fell in the slit as a function of wavelength, and adjusting the calibrated spectrum accordingly. 

For readers who are interested in making use of the Aug. 17-18 MagE spectrum, we urge caution in interpreting the data at wavelengths redder than $\sim7000$~\AA.  As a specific example, the apparent step in the spectrum at $\sim7200$~\AA\ is not a real feature.  It occurs at the breakpoint between two spectral orders and at a wavelength where there is significant telluric absorption, which makes matching the continuum levels of the neighboring orders difficult. 

The Aug. 18-19 spectra were obtained with the slit 22\arcdeg\ away from parallactic angle, causing a loss of flux at short wavelengths, which we corrected as described above for LDSS-3.

The statistical uncertainties on the MagE spectra were calculated by standard error propagation.  We added these in quadrature with additional uncertainties based on the seeing losses, telluric absorption corrections, and the overlap between adjacent spectral orders.  To be conservative we applied generous uncertainties for each of these effects.  For wavelengths within 20~\AA\ of where orders overlap we assumed a 30\%\ uncertainty on the measured fluxes.  We also assumed that the seeing loss and telluric corrections each had an uncertainty of 30\%.

\subsection{IMACS Observations}

We observed SSS17a with IMACS \cite{Dressler2006} on the Magellan/Baade telescope approximately 48 hours after its discovery on 2017 Aug. 19-20 using the f/2 camera and the 300 lines/mm grism at a blaze angle of $17.5\arcdeg$.  The spectrum was obtained through a 0.9~arcsec-wide slit providing a spectral resolving power of $R\sim1000$. We reduced and extracted the IMACS spectrum using standard routines in IRAF, including a telluric correction.

\subsection{MIKE Observations}

We obtained spectra of SSS17a totaling 1.03~hours of integration time with the MIKE spectrograph \cite{Bernstein2003} using a $2~{\rm arcsec} \times 5~{\rm arcsec}$ slit beginning at UT 00:18 on 2017 Aug 19.  The wide slit ensured that we captured as much light as possible from the fading transient.  For light that fills a 2~arcsec slit the resulting resolving power is $R=13000$ in the red ($\lambda>5000$~\AA) and $R=16000$ in the blue ($\lambda<5000$~\AA).  However, the seeing of $\sim1$~arcsec during the observations provides resolving power a factor of $\sim2$ higher for the SSS17a spectrum.  We reduced these data using the Carnegie Python pipeline \cite{Kelson2003}.  The spectrum has a signal-to-noise ratio of 8 per pixel at 5900\,\AA\ and 12 per pixel at 6600\,\AA\, and is shown in Figure S1.

\subsection{Calibrating and Dereddening Spectra of SSS17a}

We futher calibrate the non-MIKE spectra (LDSS-3, MagE, and IMACS) against photometric measurements of SSS17a by extracting synthetic photometric magnitudes for each filter that was completely contained in the wavelength range covered by the spectrum and for which we could either interpolate the photometric light curves \cite{Coulter2017, Drout17} or extrapolate them by no more than 1 hour.  The exception is the final (8.46 day) spectrum, where the $i$ band magnitude was linearly extrapolated by 1 day. Then we determined the best-fitting line to the difference between the observed and synthetic photometry as a function of central wavelength, and scaled each spectrum by this fit.  The VPH-blue and VPH-red observations on the first night only covered the wavelengths of $g$ and $i$, respectively. Thus, for these two spectra we could not correct any wavelength-dependent flux calibration issues and instead only applied a zero-point correction. The bands used to calibrate each spectrum are listed in Table~S1. 

We adopt reddening and extinction estimates of $E(B-V)$ = 0.106 and $A_{\rm V} = 0.34$\,mag for our analysis based on the far-infrared dust maps of \cite{Schlafly2011}.  These measurements are consistent with the extinction value of $A_{\rm V} = 0.37 \pm 0.06$\,mag determined by \cite{Green2015} using Pan-STARRS1 stellar colors.  As a consistency check, we also estimated the extinction along the line of sight to SSS17a with the Na\,I~D absorption lines in the MIKE data. We modeled the Na lines from the Milky Way with a single Gaussian component for each of the D2 and D1 lines, which provides an accurate fit to the spectrum (Figure~S1). The spectrum can also be fit with additional weaker components, but because of the modest signal-to-noise ratio of the data and the presence of residuals from the subtraction of the telluric Na~D emission lines at nearly the same velocity, those features are not statistically significant.  We measured a best-fitting heliocentric velocity of $4.7 \pm 1.2$\,km\,s$^{-1}$ for this absorbing gas, with a full width at half maximum (FWHM) of $26 \pm 3$\,km\,s$^{-1}$ and equivalent widths (EWs) of $328 \pm 52$\,m\AA\ for the D2 line and $256 \pm 62$\,m\AA\ for the D1 line.  We calculated a Na\,I column density of $5.2 \times 10^{12}$\,cm$^{-2}$ from the EW measurements using a standard curve-of-growth analysis. This column density corresponds to a V-band extinction $A_{\rm V} \approx 0.4$\,mag \cite{Phillips2013}.  Although this value is consistent with the results of \cite{Schlafly2011} and \cite{Green2015}, because of the uncertainties involved in translating Na~D absorption strength into extinction we do not make further use of it in this paper.  

We correct the observed spectra for this foreground reddening using a Milky Way extinction curve \cite{Cardelli1989}. The extinction curve is parameterized by the value $R_V \equiv A_V/E(B-V)$, where $E(B-V) \equiv A_{B} - A_{V}$ is the selective extinction between the $B$ and $V$ photometric bands ($A_B$ is the total extinction in the $B$-band). We assume $R_V = 3.1$.

\subsection{Synthetic Photometry of SSS17a}

We measured synthetic photometry in any Sloan ($griz$) or Johnson/Cousins (BVRI) photometric bandpass whose transmission window falls within the wavelength range of the observed spectrum.  In order to facilitate direct comparison to broadband observations of SSS17a, this photometry was performed after correction for slit losses, but before the correction for Milky Way reddening described above.  To estimate the uncertainties on the synthetic magnitude measurements we run a Monte Carlo simulation.  We randomly redraw the photometric measurements 50,000 times based on the observed photometry \cite{Drout17}, assuming that the photometric uncertainties are normally distributed around the measured magnitudes with a width given by the photometric uncertainties.  For each draw, we recalibrate the spectrum to the drawn photometry and then perform synthetic photometry. We adopt the 16th and 84th percentiles of the distribution of synthetic magnitude measurements in the Monte Carlo simulation as the $1\sigma$ uncertainties on each measurement.  For the day 8.46 spectrum we do not make any synthetic measurements because the photometry used to calibrate that spectrum was extrapolated from one day earlier.  We report the synthetic photometry and associated uncertainties in Table~S2. 

\subsection{GRB130603B Spectra}

Both GRB130603B spectra plotted in Figure~2 were presented in \cite{deUgartePostigo2014}. The GTC/OSIRIS spectrum was taken from that paper while we retrieved and reduced the raw X-Shooter optical and near-infrared spectra using the ESO Reflex environment and the X-Shooter standard pipeline recipes. These spectra are mostly featureless within the noise, except for telluric lines and narrow Mg and Ca absorption.

\section{Constraints on Host Galaxy Absorption Lines}

We searched the MIKE and MagE spectra for host galaxy absorption or emission lines.  The host galaxy, NGC~4993 \cite{Coulter2017}, has a redshift of $z=0.00988$ \cite{Pan17}.  We are unable to detect any absorption lines associated with the host galaxy.This result is in agreement with the results of VLT/X-Shooter observations \cite{Pian2017}.  Specifically, we do not detect host galaxy Na~D absorption, from which we conclude that all of the extinction along the line of sight to SSS17a is located in the Milky Way.  
Assuming the same linewidth as for the Milky Way Na~D absorption (S1.5) and using the formula given by \cite{Frebel2006}, we place a 2$\sigma$ upper limit on host galaxy Na~D absorption of $92$~m\AA\ (for either the D2 or D1 lines).  In comparison, the detected Na~D EWs for GRB130603B were $530 \pm 90$~m\AA\ and $590 \pm 80$~m\AA\ for the D2 and D1 lines, respectively \cite{deUgartePostigo2014}.  The signal-to-noise ratio of the MIKE spectrum in the blue is too low to place meaningful constraints on Ca~H and K absorption.  
The lack of any neutral gas in NGC~4993 near the site of the merger might indicate that SSS17a was on the near side of the host galaxy, or that NGC~4993 is a largely gas-free system.  We also do not detect H$\alpha$ in either absorption or emission (with a 2$\sigma$ upper limit of 63~m\AA\ for a linewidth of 25\,km\,s$^{-1}$) or the O\,III~$\lambda5007$\,\AA\ emission line, although the signal-to-noise ratio of the spectrum near 5000\,\AA\ is low because it is close to the wavelength of the dichroic of the spectrograph.

\section{BlackBody Spectral Fitting}

We fit blackbody models to the observed dereddened rest-frame spectra from the first night using Markov Chain Monte Carlo (MCMC) methods.  For a single MCMC run, the statistical uncertainties described above for the LDSS-3 (S1.1) and MagE (S1.2) spectra define the width of the posterior probability distributions for the blackbody temperatures and radii.  We carry out additional Monte Carlo simulations to incorporate the effects of systematic uncertainties on the spectra and the photometric flux measurements to which they are tied (S1.5).  For the LDSS-3 spectrum we draw 500 random positions of the source within the slit from a Gaussian distribution with a FWHM of 0.1~arcsec.  We scale the observed spectrum by the slit losses from differential atmospheric refraction and seeing for each of the randomly drawn source positions to create 500 Monte Carlo spectra.  We also randomly draw 5 $g$ and $i$ magnitudes for each source position based on the measured photometric uncertainties and scale the Monte Carlo spectra accordingly.  We then re-run the MCMC with the resulting spectra.  The reported uncertainties on the BB temperatures and radii are the 90\%\ confidence intervals from these 2500 Monte Carlo iterations.  Because the MagE spectrum was obtained at parallactic angle there is no systematic component to the spectroscopic uncertainties.  The uncertainties related to the photometry still apply, so we randomly draw 2000 $g$ and $i$ magnitudes, scale the spectrum, re-run the MCMC on the Monte Carlo spectra, and define the uncertainties as above.

Figure~1 shows the observed spectra as well as shaded regions with lower and upper bounds corresponding to the blackbody spectra expected for the 5th and 95th percentile temperature and radii obtained from the fits, respectively. 


\section{Existing Kilonova Models}

There are many theoretical models of kilonovae in the literature, but they have been developed with few observational constraints.  
To investigate whether any aspects of the SSS17a spectra agree with theoretical predictions, we searched the three main classes of published models: i) classic lanthanide-rich red kilonova \cite{Barnes2013}, ii) lanthanide-poor accretion disk winds following the merger \cite{Kasen2015}, and iii) a model including both hydrodynamical ejecta and wind from a black hole (BH)-neutron star (NS) merger \cite{Fernandez2017}.  We find that no existing model we considered simultaneously produced satisfactory fits to the spectroscopic features, color, evolution, and luminosity of our spectroscopic time series.
However, studying the ways in which models do agree often leads to additional physical insights. Therefore, we re-examined the models for qualitative similarities with the data by scaling the model luminosities at each epoch to match the observed spectroscopic times series.  Even with that additional freedom almost all models were unsatisfactory.  However, for the two models shown in Figures~4A and 4B some features resemble the spectra of SSS17a.  
The NS-NS merger model of \cite{Barnes2013} has 0.1 solar masses of ejecta composed of Ca, Fe, and Nd distributed in a broken power law density profile with $v=0.2c$.  The resulting spectra exhibit smooth red continua with a similar shape to the observed spectra from 4.51 days onward, with the exception of the predicted emission feature at $\sim6000$~\AA\ (Figure~4A).
Conversely, the disc wind outflow model of \cite{Kasen2015} with a NS with lifetime of 0 ms reproduces the spectral slopes during the first 3.5 days.  However, it strongly over-predicts absorption features and under-predicts the photospheric velocity after the merger.

\clearpage

\vspace{0.2in}
\includegraphics*[width=0.90\textwidth, trim=0cm 0 0 0]{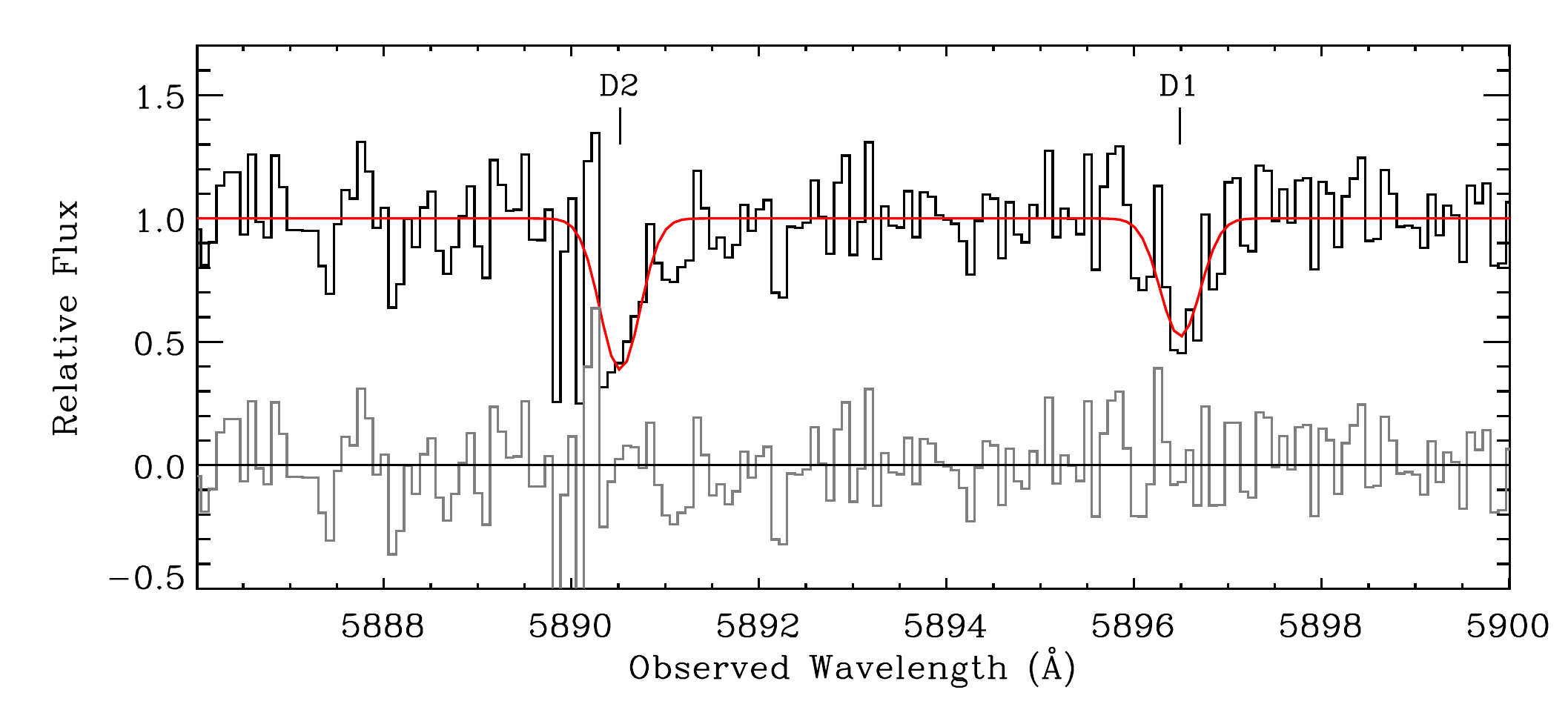}
\newline
\noindent{\bf Figure~S1.}  {{\bf High-resolution MIKE spectrum of SSS17a.} The plotted wavelength range is centered on the Na\,I~D absorption lines from the interstellar medium of the Milky Way.  The red curve is our Gaussian fit to the spectrum, and the gray spectrum below is the residuals from the fit.}

\clearpage

\begin{table}
\begin{center}
\caption{{\bf Spectroscopic Observations of SSS17a.} Columns include the UT Date and Julian Date (JD) at the start of the observation, rest frame days since the gravitational wave trigger (phase), the telescope and instrument with which the observations were taken, the rest frame wavelength range of the spectrum, the total exposure time of the spectrum, and the broad-band photometric filters that were used to calibrate the spectrum. }\scriptsize
\begin{tabular}{@{}lcrlcrc}\hline\hline
UT Date & JD & phase & Telescope/Instrument & Wavelength range & Exposure & Calibrating \\   &   & (days)  &   & (\AA) & (s) & Filters \\ \hline

2017-08-18 00:26:16.9  & 2457983.518251  & 0.49  & Magellan-Clay/LDSS-3   & $3780\--10200$  & 300   & $g,i$ \\
2017-08-18 00:40:08.6  & 2457983.527877  & 0.50  & Magellan-Clay/LDSS-3   & $3800-6200$  & 300   & $g$ \\ 
2017-08-18 00:52:08.6  & 2457983.536204  & 0.51  & Magellan-Clay/LDSS-3   & $6450\--10000$  & 600   & $i$ \\ 
2017-08-18 01:26:22    & 2457983.559977  & 0.53  & Magellan-Baade/MagE   & $3650\--10100$  & 322   & $g,i$ \\ 
2017-08-18 23:47:36.7  & 2457984.491397  & 1.46  & Magellan-Clay/LDSS-3   & $3820\--9120$   & 600   & $g,V,r,i$ \\
2017-08-19 00:18:11    & 2457984.512627  & 1.48  & Magellan-Clay/MIKE    & $3900\--9400$  & 3710   & N/A \\ 
2017-08-19 00:35:25    & 2457984.524595  & 1.50  & Magellan-Baade/MagE   & $3800\--10300$  & 3000   & $g,V,r,i$ \\ 
2017-08-20 00:26:27.9  & 2457985.518368  & 2.49  & Magellan-Baade/IMACS  & $4355\--8750$   & 2100  & $V,r,i$ \\
2017-08-20 23:45:52.7  & 2457986.490193  & 3.46  & Magellan-Clay/LDSS-3   & $4450\--10400$  & 3000   & $V,r,i,z$ \\
2017-08-22 00:50:33.7  & 2457987.535112  & 4.51  & Magellan-Clay/LDSS-3   & $5010\--10200$  & 3600   & $i,z$ \\
2017-08-24 23:33:51.5  & 2457990.481846  & 7.45  & Magellan-Clay/LDSS-3   & $6380\--10500$  & 3600   & $i,z$ \\
2017-08-25 23:39:18.1  & 2457991.485626  & 8.46  & Magellan-Clay/LDSS-3   & $6380\--10500$  & 3600   & $i^{*},z$ \\
\hline
\end{tabular}\label{tab:Observations}
\end{center}
{$^{*}$This data point was extrapolated from the latest measurement on the previous night.}
\end{table}

\begin{table}
\begin{center}
\caption{{\bf Synthetic photometry measurements from spectroscopic observations of SSS17a.} Columns include the Julian Date (JD) of the spectroscopic observation and the broadband filter of the measurement ($grizBVR$ or $I$). Synthetic photometry was performed prior to correction for foreground Milky Way reddening. $griz-$band magnitudes are presented on the AB magnitude scale. $BVRI-$band magnitudes are presented on the Vega magnitude scale. $1\sigma$ uncertainties are listed. }\scriptsize
\begin{tabular}{@{}l|llll|llll}\hline\hline
JD  & $g$  & $r$ & $i$ & $z$ & $B$ & $V$ & $R$ & $I$ \\ & \multicolumn{4}{c|}{(AB mag)} & \multicolumn{4}{c}{(Vega mag)} \\
\hline

2457983.518251 &  $17.39^{+0.02}_{-0.02}$ &  $17.33^{+0.02}_{-0.01}$ &  $17.47^{+0.02}_{-0.02}$ &  $17.67^{+0.03}_{-0.03}$ &  ... &  $17.33^{+0.02}_{-0.02}$ &  $17.15^{+0.01}_{-0.01}$ &  $17.08^{+0.02}_{-0.02}$ \\
2457983.527877 &  $17.41^{+0.02}_{-0.02}$ &  ... &  ... &  ... &  ... &  ... &  ... &  ... \\
2457983.559977 &  $17.45^{+0.02}_{-0.02}$ &  $17.34^{+0.01}_{-0.01}$ &  $17.51^{+0.02}_{-0.02}$ &  $17.59^{+0.03}_{-0.02}$ &  $17.63^{+0.03}_{-0.02}$ &  $17.35^{+0.02}_{-0.02}$ &  $17.17^{+0.01}_{-0.01}$ &  $17.12^{+0.02}_{-0.02}$ \\
2457984.491397 &  $18.64^{+0.02}_{-0.02}$ &  $17.92^{+0.01}_{-0.01}$ &  $17.78^{+0.02}_{-0.02}$ &  ... &  ... &  $18.18^{+0.01}_{-0.01}$ &  $17.67^{+0.01}_{-0.01}$ &  $17.31^{+0.03}_{-0.02}$ \\
2457985.518368 &  ... &  $18.98^{+0.04}_{-0.05}$ &  $18.36^{+0.02}_{-0.02}$ &  ... &  ... &  $19.49^{+0.07}_{-0.08}$ &  $18.57^{+0.03}_{-0.03}$ &  $17.82^{+0.04}_{-0.04}$ \\
2457986.490193 &  ... &  $19.88^{+0.05}_{-0.05}$ &  $18.91^{+0.04}_{-0.04}$ &  $18.38^{+0.05}_{-0.05}$ &  ... &  $20.41^{+0.08}_{-0.08}$ &  $19.35^{+0.04}_{-0.04}$ &  $18.26^{+0.04}_{-0.04}$ \\
2457987.535112 &  ... &  $21.00^{+0.14}_{-0.12}$ &  $19.43^{+0.04}_{-0.04}$ &  $18.75^{+0.08}_{-0.08}$ &  ... &  $21.83^{+0.22}_{-0.17}$ &  $20.19^{+0.09}_{-0.07}$ &  $18.67^{+0.04}_{-0.04}$ \\
2457990.481846 &  ... &  ... &  $21.44^{+0.20}_{-0.20}$ &  $19.91^{+0.07}_{-0.07}$ &  ... &  ... &  ... &  $20.37^{+0.11}_{-0.10}$ \\

\hline
\end{tabular}\label{tab:synphot}
\end{center}
\end{table}

\clearpage

\end{document}